# Broadband millimeter-wave frequency mixer based on thin-film lithium niobate photonics

Xiangzhi Xie[1,†,*], Hanke Feng[1,†], Yuansheng Tao[1], Yiwen Zhang[1], Yikun Chen[1], Ke Zhang[1], Zhaoxi Chen[1] & Cheng Wang[1,*]

[1] *Department of Electrical Engineering & State Key Laboratory of Terahertz and Millimeter Waves, City University of Hong Kong, Kowloon, Hong Kong, China*

*Corresponding authors: XZ.Xie@cityu.edu.hk; cwang257@cityu.edu.hk

†These authors contributed equally to this work.

**Abstract** —Frequency mixers are fundamental components in modern wireless communication and radar systems, responsible for up- and down-conversion of target radio-frequency (RF) signals. Recently, photonic-assisted RF mixers have shown unique advantages over traditional electronic counterparts, including broad operational bandwidth, flat frequency response, and immunity to electromagnetic interference. However, current integrated photonic mixers face significant challenges in achieving efficient conversion at high frequencies, especially in millimeter-wave bands, due to the limitations of existing electro-optic (EO) modulators. Additionally, high-frequency local oscillators in the millimeter-wave range are often difficult to obtain and expensive, leading to unsatisfactory cost and restricted operational bandwidth in practice. In this paper, we harness the exceptional EO property and scalability of thin-film lithium niobate (TFLN) photonic platform to implement a high-performance harmonic reconfigurable millimeter-wave mixer. The TFLN photonic circuit integrates a broadband EO modulator that allows for extensive frequency coverage, and an EO frequency comb source that significantly reduces the required carrier frequency of the local oscillator. We experimentally demonstrate fully reconfigurable frequency down-conversion across a broad operational bandwidth ranging from 20 GHz to 67 GHz, with a large intermediate frequency of 20 GHz, as well as up-conversion to frequencies of up to 110 GHz. Our integrated photonic mixing system shows dramatically improved bandwidth performance, along with competitive indicators of frequency conversion efficiency and spurious suppression ratio, positioning it as a promising solution for future millimeter-wave transceivers in next-generation communication and sensing systems.

## I. Introduction

Frequency mixers (FMs) are fundamental components in modern communication systems, enabling the conversion between lower intermediate frequencies (IF) and higher radio frequencies (RF) during signal transmission and reception [1]. With the rapid development of high-speed communication systems [2,3], high-resolution multifunctional radar [4–6], and software-defined payloads [7], higher center frequencies are increasingly utilized to free up spectrum resources, enabling broadband communication with ultrahigh transmission capacity and sensing systems with improved resolution. Compared with traditional electronic FMs, photonics-based FMs have recently demonstrated unique advantages, including large operational frequency range, broad IF bandwidth, and small frequency-dependent loss [8]. Considerable efforts have been directed toward the implementation of photonic-based frequency conversion systems utilizing nonlinearities in semiconductor optical amplifiers [9–12] and electro-optic (EO) modulators [13–19]. However, most demonstrations above rely on discrete optoelectronic components, resulting in unsatisfactory system size, complexity, and cost.

The rapid development of integrated photonic circuits provides an effective method to reduce the size, cost, and power consumption of microwave photonics systems [20,21]. A variety of FMs based on integrated photonic platform have been proposed and experimentally demonstrated with compact footprints [22–30]. Typically, two parallel phase modulators (PMs) are used to encode the input and local oscillator (LO) signals respectively, before the modulated signals are filtered and beat at a photodetector to generate mixed signals. For example, a fully packaged integrated photonic mixer has been developed, incorporating an indium phosphide (InP) laser chip and a silicon-on-insulator (SOI) photonic circuit with two PMs, with an operational bandwidth of 20 GHz [23]. A similar concept has also been explored using bulk lithium niobate dual-parallel modulators, however only an operational bandwidth frequency range of 16 GHz has been achieved due to the limited bandwidth performance of the modulators [26]. To date, it has remained a significant challenge for integrated photonic FMs to achieve high operational frequency into the millimeter-wave bands and broad IF bandwidths simultaneously. Apart from bandwidth limitations of EO modulators, challenges also come from the LOs, which become substantially more costly and difficult to obtain at high carrier frequencies. Optical frequency combs (OFCs) could provide an intriguing solution to this, as their coherent and equally spaced harmonics could be used as calibrated frequency rulers for photonic LOs [31]. Kerr optical micro-combs have been utilized as photonic LOs for mixing RF signals up to 40 GHz; however, the EO modulation components have remained off-chip due to incompatibility between EO and Kerr photonic platforms [28].

Thin-film lithium niobate (TFLN) is an emerging photonic platform to address these challenges. Its strong electro-optic (Pockels) effect enables not only efficient and high-speed EO modulators [32–40], but also broadband electro-optic frequency comb sources [41–46]. In contrast to Kerr combs [47–49], EO combs based on cascaded modulators allow accurate and flexible control of comb spacing by changing the RF driving frequency, and can be engineered to generate near-flat amplitude distribution covering a wide spectral range [50,51]. More importantly, the TFLN platform also features low optical propagation loss [52] and wafer-scale manufacturability, allowing for the monolithic integration of various device building blocks for high-performance system-level microwave photonic applications on chip scales [6,53]. This has opened up the possibility to integrate OFC generator [50], broadband EO modulator [35], and high-performance optical filters [34] on the same TFLN chip for efficient and broadband millimeter-wave mixing.

In this work, we fulfill this promise and demonstrate both frequency down-conversion from and up-conversion into millimeter-wave frequencies with ultrabroad IF bandwidths using a single TFLN photonic chip. Specifically, an on-chip EO comb generator produces a series of LOs that cover a broad spectral range with a tunable repetition rate. A high quality ($Q$) factor add-drop ring resonator is used to extract one target comb line as the LO for a particular mixing process. Thanks to the widely distributed spectral references, the frequency requirements of the LO have been significantly reduced (in our case ~ 20 GHz). A Mach-Zehnder Modulator (MZM), with a bandwidth exceeding 67 GHz, is utilized to achieve EO conversion of input signal under carrier-suppressed conditions. The modulated input signal beats with the selected LO at a photodetector (PD) to complete the mixing process. Experimentally, we leverage three different LOs to demonstrate RF down-conversion with a broad operational bandwidth ranging from 20 GHz to 67 GHz and a high IF bandwidth of 20 GHz. To date, this represents the widest RF and IF bandwidth in reported photonic RF mixers. Additionally, the conversion efficiency (CE) and spurious suppression ratio (SSR) exhibit competitive performance across the board. We further demonstrate RF signal up-conversion to 40 GHz and 110 GHz bands by utilizing the 2nd and 5th order harmonics of the EO comb, respectively. Our compact broadband photonic RF mixer, with significantly reduced system size, complexity, and potential costs, is a promising solution for future millimeter-wave transceivers in next-generation communication and sensing systems.

## II. Experimental Results

The schematic illustration of the TFLN FM system is shown in Fig. 1, with insets (i-vii) demonstrating the optical and electrical spectra at various locations. A continuous-wave (CW) laser is edge-coupled to the TFLN chip as the optical carrier, $f_0$ (i) and split into two paths. Carrier signal in the top path is phase-modulated by an RF LO signal at ~ 20 GHz (iv) through a double-pass EO PM, enabling OFC generation (ii). The phase modulation process is enhanced by allowing optical signals to pass through the electrodes for two round trips with the assistance of a low-loss and low-crosstalk waveguide crossing [54], which reduces electrical power consumption by a factor of four [55]. An add-drop ring resonator follows, with its center wavelength widely tunable through the thermal-optic effect. This allows both upper and lower sidebands with significant spectral spacing from carrier to be selected as photonic LOs during the frequency mixing process, effectively realizing LOs at frequency multiples of the input RF LO frequency. The target RF signal (v) is modulated in the lower path under carrier-suppressed conditions through a broadband MZM, which is biased at the null point by applying a DC voltage. In the down-conversion process, the upper sideband of the OFC that is close to the target signal is filtered out (iii) and combined with the optically modulated RF signal (vi) before being sent to the PD. The beating process in the PD converts the high-frequency RF signal to an IF signal. For example, the $n$th order harmonic of the OFC with frequency $f_0 + nf_{LO}$ is mixed with the target optical signal at frequency $f_0 + f_{RF}$, resulting in a beat note that is shifted to a lower frequency band with a center frequency of $f_{RF} - nf_{LO}$. During frequency up-conversion, the lower sideband with frequency $f_{RF} - nf_{LO}$, which is distant from both the target signal and the carrier frequency, beats with the target signal at low frequency, enabling a dramatic increase in the center frequency. The input optical spectrum and output electrical spectrum of the PD are illustrated in (vii) and (viii), respectively.

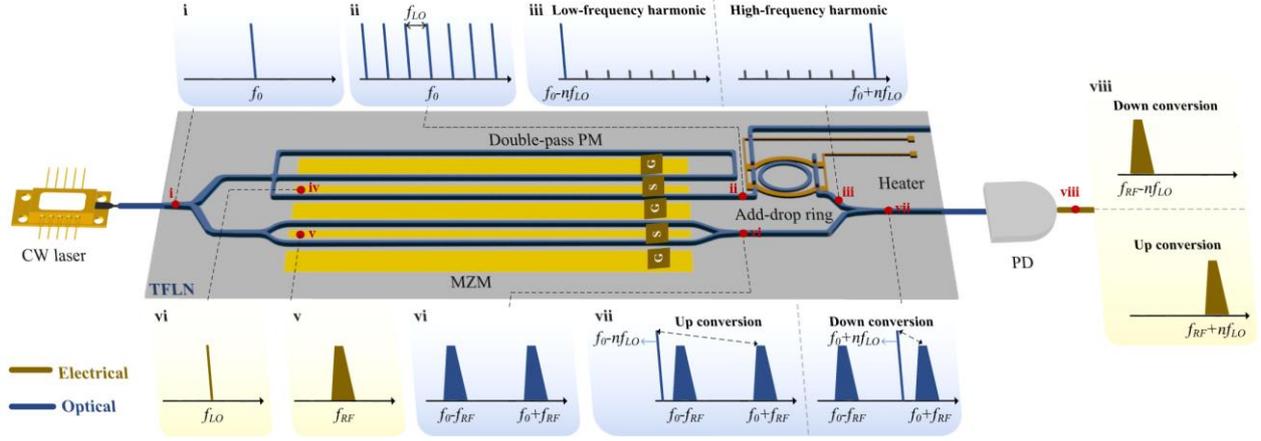

**Figure 1** Schematic illustration and working principle of the TFLN frequency mixing (FM) system. A double-pass phase modulator (PM) generates electro-optic (EO) frequency combs to serve as a series of equally-spaced and reconfigurable photonic local oscillators, one of which is selected by an add-drop ring resonator. A broadband Mach-Zehnder modulator (MZM) biased at the null point provides broadband conversion of target RF signals into optical domain. Frequency down-conversion and up-conversion over a broad operational bandwidth is achieved by tuning the center wavelength of the add-drop filter to select a particular LO signal that are close to or far from $f_0 + f_{RF}$ and beating it with the optically modulated signal at a photodetector (PD). Insets (i-vii) illustrate the optical and electrical spectra at various points on the chip, with optical signals shown in blue and electrical signals in yellow. CW laser: continuous-wave laser.

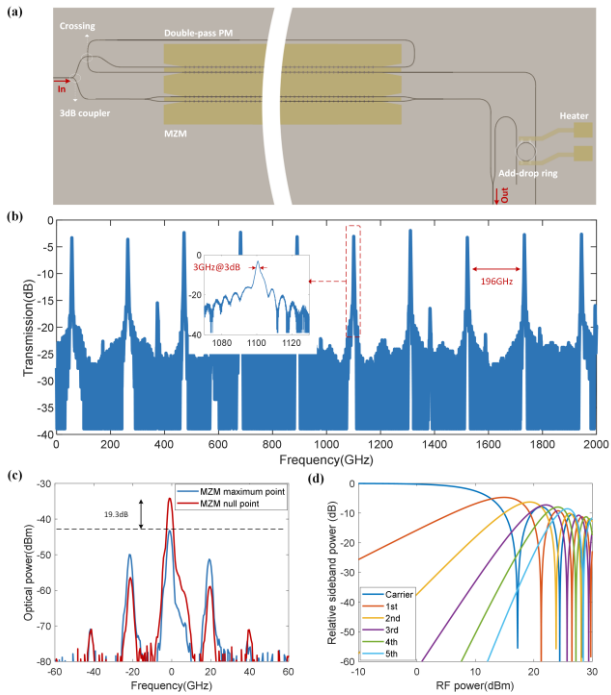

**Figure 2** Characteristics of the proposed TFLN FM system. (a) Optical image (false-color) of the FM system. (b) Frequency characteristics of the add-drop ring filter. (c) Small-signal double-sideband (DSB) modulation of the MZM showing a carrier suppression ratio of 19.3 dB. (d) Numerically simulated optical sideband powers as functions of input RF power.

The TFLN FM system is fabricated using a 4-inch wafer-scale manufacturing process using an ultraviolet (UV) stepper lithography system (see Methods). Fig. 2(a) shows the false-color optical image of the FM system, consisting of a double-pass phase modulator, an add-drop ring with a high extinction ratio, and a push-pull MZM. The optical transmission spectrum of the add-drop ring is shown in Fig. 2(b), featuring a free spectral range (FSR) of 196 GHz and a 3 dB bandwidth of the transmission peak ~ 3 GHz. The MZM exhibits a half-wave voltage of approximately 3 V, and an EO bandwidth exceeding 67 GHz, facilitated by advanced slotted electrodes [56]. Figure 2(c) shows double-sideband (DSB) modulation spectra of the MZM under small-signal carrier-unsuppressed (red) and carrier-suppressed (blue) situations, revealing a carrier suppression ratio of 19.3 dB. Figure 2(d) shows the numerically simulated optical powers of different sideband orders of the PM as functions of RF-driving power, showing the optical power of each sideband can be redistributed by varying the input RF power. Assuming a PM half-wave voltage of 3 V, the optical power of the 1st to 5th order sidebands reaches a maximum at input RF powers of 16.4 dBm, 19.8 dBm, 22.3 dBm, 24.3 dBm, and 26.2 dBm, respectively.

The experimental results of OFC generation and LO selection are shown in Fig. 3(a) and 3(b), respectively. The RF frequency is set to 20 GHz, with the driving power varying from 10.2 dBm to 22.9 dBm, resulting in a gradual expansion of the spectral range. The optical power of the target sideband is optimized to maximize the extinction ratio between the adjacent mode by carefully tuning the RF driving power, which closely aligns with the numerical simulation results presented in Fig. 2(d). Carefully tuning the add-drop ring allows accurate selection of the target sideband with excellent mode suppression. The measured extinction ratios between filtered upper adjacent line and the target signal are 37.0 dB, 35.4 dB, and 33.6 dB at 20 GHz, 40 GHz, and 60 GHz, respectively, as shown in Fig. 3(b). In a down-conversion process, the lower adjacent lines lead to mixing signals at much higher frequencies (20 GHz plus target mixed frequency), which can be naturally filtered out using a PD with modest bandwidth, and therefore are less problematic in practical applications. The measured mode suppression ratio slightly deteriorates as the sideband order increases, primarily due to the reduced power difference with the adjacent sidebands. The photonic harmonic generation process effectively multiplies the frequency of the RF LO, providing a reconfigurable LO with a broadband tuning range up to 60 GHz for the mixing process.

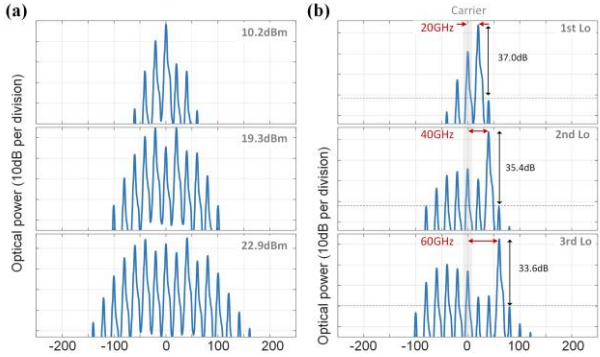

**Figure 3** The generated multi-stage photonic LO signal. (a) The optical spectrum of the OFC at different RF drive powers. (b) The selected 1st to 3h sidebands after passing through the add-drop ring.

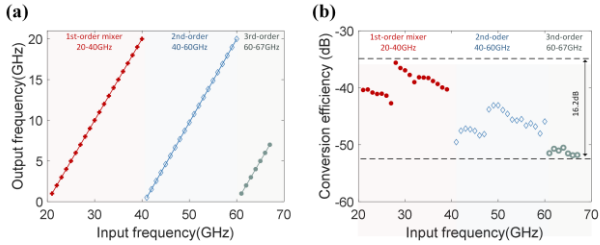

**Figure 4** Frequency down-conversion performance from 20 GHz to 67 GHz. (a) Frequency mapping of input and down-converted signals. (b) Conversion efficiencies at various input frequencies.

Figure 4 shows the experimental results of the down-conversion processes from a broad range of frequencies from 20 GHz to 67 GHz using three effective LOs with a frequency interval of 20 GHz. The mapping relationship between input and output frequencies is illustrated in Fig. 4(a), where the IF bandwidth is 20 GHz and the operational bandwidth is 67 GHz. The first, second, and third order LOs cover the ranges of 20 - 40 GHz, 40 - 60 GHz, and 60 - 67 GHz, respectively. The dots represent measurement results from 20 GHz to 67 GHz with a step of 1 GHz, while the lines indicate linear fitting results. The operational bandwidth of the down-converter is primarily limited by the available microwave source and RF probe, which could potentially be further extended deep into the millimeter-wave or even terahertz frequency ranges leveraging ultrabroad-band TFLN modulators [57]. The measured down-conversion efficiency is shown in Fig. 4(b), where the average value is –44.1 dB without the use of electrical amplifiers, which is on par with typical passive electronic RF mixers but with much larger operational and IF bandwidths [58,59]. The flatness of conversion efficiency over the measured frequency range is 16.2 dB. The conversion efficiency slightly decreases at increased frequencies, likely due to performance roll-off of the MZM and the OFC generator.

To evaluate the spurious suppression performance of our photonic FM, we measure the output RF spectra over a broad frequency range at input frequencies of 21GHz, 41 GHz, and 61 GHz, as shown in Fig. 5(a), (b) and (c), respectively. Taking the first order down-converter as an example, the input frequency of 21 GHz is converted to 1

GHz after beating with the first order LO, with the most prominent spurious frequency located at $2f_{LO} - f_{RF}$, i.e. 19 GHz. The measured spurious suppression ratio (SSR) in this case is 37.2dB. Similarly, the measured SSR of the second order mixing process is 29.3dB at $3f_{LO} - f_{RF}$, and that of the third order mixing process is 26.1dB at $4f_{LO} - f_{RF}$, both of which are located at 19 GHz in these experimental settings.

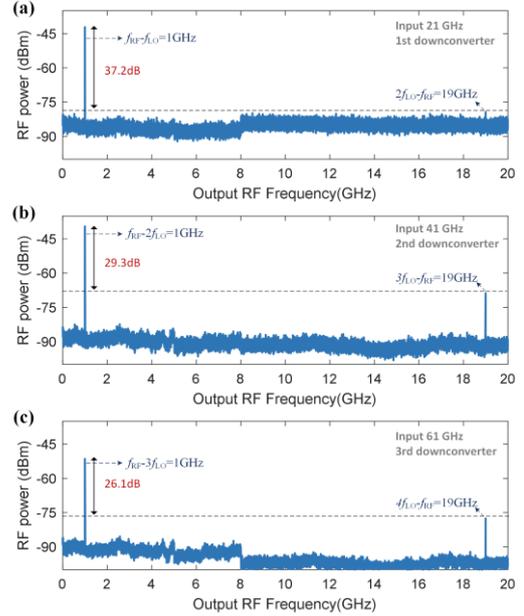

**Figure 5** Down-converted RF spectra corresponding to input frequencies of (a) 21 GHz, (b) 41 GHz, and (c) 61 GHz.

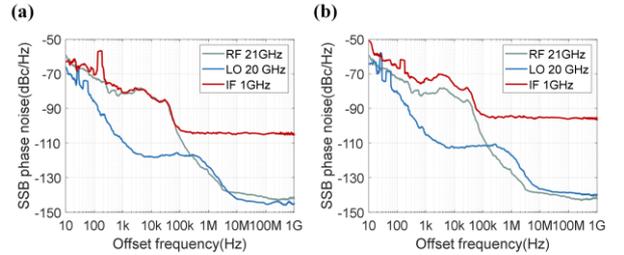

**Figure 6** Single sideband (SSB) phase noise of the (a) first- and (b) second-order mixing processes.

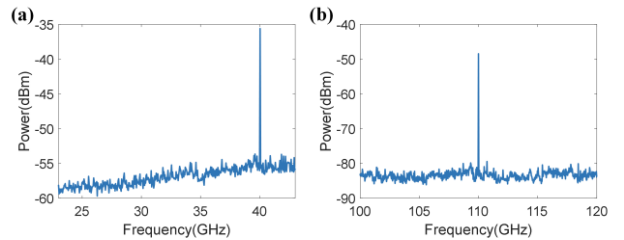

**Figure 7** Millimeter-wave signal generation at (a) 40 GHz and (b) 110 GHz enabled by frequency up-conversion.

We further characterize the single sideband (SSB) phase noises of the mixed signals, as shown in Fig. 6. The phase noises of the input LO, input RF signal, and output IF signal are represented by blue, cyan, and red lines, respectively. Within the offset range of 100 kHz, the SSB phase noise of the IF signal generally follows the trend of the input RF signal. As the offset frequency increases, the

SSB phase noise remains relatively unchanged (-104.3 dBc/Hz for first-order mixer and -95.2 dBc/Hz for second-order mixer), likely influenced by the noise figures of the optical link. The measured IF SSB noise deteriorates slightly in higher-order mixers, which can be attributed to cascaded phase noise during the harmonic generation process.

We finally demonstrate high-frequency millimeter wave generation at up to 110 GHz using the frequency up-conversion process, as shown in Fig. 7. As a proof of concept, we bias the MZM at the maximum transfer point without signal modulation. The optical carrier is directly mixed with one selected harmonic frequency of the OFC. The 40 GHz RF signal in Fig. 7(a) is generated by frequency doubling the LO frequency set to 20 GHz, resulting in an output RF power of -36 dBm without amplification. The generated millimeter-wave power does not significantly vary at different frequencies, since the optical power of each sideband is not substantially affected after optical amplification with the same optical saturation power of the PD. We further generate THz signals at 110 GHz through a fivefold multiplication process [Fig. 7(b)]. Considering the analog bandwidth of the available THz receiver, the LO frequency is slightly adjusted to 22 GHz. The THz receiver consists of an electrical down-converter and an electrical spectrum analyzer, with a total conversion loss of approximately 5 dB. Thus, the actual output power of the THz signal is around -43 dBm.

## III. Discussion

Table 1 provides a performance comparison of reported integrated photonic FMs and this work. Our TFLN solution shows excellent operational bandwidth performance for both up- and down-conversion deep into millimeter wave bands, representing the widest RF and IF bandwidths to date. Our FMs also feature competitive conversion efficiency (CE) and SSR among other schemes with relatively flat CEs over the entire frequency range. The operational bandwidth has not been fully explored due to the cut-off frequency of the RF probe, with could be further extended into terahertz frequencies leveraging our recently demonstrated TFLN modulators with effective EO modulation bandwidths up to 500 GHz.

For the up-conversion part, we here only measured frequency-doubled and fivefold-multipled signals due to limitations in our measurement instruments. The spurious harmonics with a frequency interval of 20 GHz is unlikely to significantly affect applications, as they are separated far enough away and can be easily filtered out at the receiver end. The IF bandwidth of the proposed scheme is also tunable by adjusting the repetition rate of the OFC, and higher photonic harmonics can be generated by increasing the driving RF power. The OFC based on the TFLN platform, featuring a flat-top amplitude, has been demonstrated over a spectral range of 12 nm [31]. Our proposal has the potential to achieve frequency conversion at higher frequencies by utilizing RF probes and PDs with larger operational bandwidths.

**Table 1** Performance Comparison of Reported Integrated Photonic FMs

| Platform | Integrated Part | Function | RF BW (GHz) in (dB) | IF BW | CE (dB) | SSR (dB) |
|---|---|---|---|---|---|---|
| Silicon [22] | Ring modulator, MRR | DC | 20-40/25 | 0.01-3 | -62.5 | – |
| Hybrid [23] | InP Laser, SOI PM, MRR, PD | DC | 2-18/6 | 0.1-3 | -55 | – |
| Hybrid [24] | Laser, PD, bulk modulator and FBG [a] | DC | 10-18/– | 0.01-5 | -33.5 | – |
| SOI [25] | MZM, PD | DC | 4-18/– | 2-10 | -57 | – |
|  |  | UC | 3.4-14.1/25 | – | – | – |
| Hybrid [26] | Bulk LiNbO$_3$ PM, Si$_3$N$_4$ MRR | DC | 4-20/– | 0.2-1 | -35[b] | – |
| SOI [27] | Comb, CS-SSB Modulators, filter, PD | UC | 0.5-50/– | ≤4 | -66 | >20 |
| Silica glass [28] | Microcomb | DC | 23-40/3 | 8.9-25.9 | -35 | >40 |
| TFLN [29] | MZM | DC | –/– | – | -7.4 | – |
| Hybrid [30] | Laser, bulk PM, SOA, PD | DC | 0.2-0.5 | 0-0.3 | -60 | – |
| **TFLN** | **Comb, MZM, MRR** | **DC** | **20-67/5** | **0-20** | **-44.1** | **>26.1** |
|  |  | UC | 40,110/10 | – | -55.3 | – |

[a] These components are interconnected using free space micro-optics.
[b] Calculated value without the electrical amplifier
DC: down conversion, UC: Up conversion, BW: bandwidth, CE: conversion efficiency, SSR: spurious suppression ratio, SOI: Silicon-on-insulator, InP: Indium phosphide, FBG: fiber Bragg grating, MRR: microring resonator.

## IV. Conclusion

In conclusion, we demonstrate a compact photonic-assisted FM based on the TFLN platform with an extended operational bandwidth into the millimeter wave range. Experimentally, we have demonstrated a down-converter covering the range of 20 GHz to 67 GHz, as well as 110 GHz generation enabled by the up-converter. This technique shows promise for improving the RF and IF bandwidth of FMs, while significantly relaxing the requirements for the carrier frequency of the LO. These features are advantageous for applications in next-generation communication and sensing systems.

**Method**

**Design and fabrication of the devices**

Devices are fabricated from a commercially available x-cut LNOI wafer (NANOLN), featuring a 500-nm LN thin film, a 4.7-μm buried SiO$_2$ layer, and a 500-μm silicon substrate. First, a layer of SiO$_2$ is deposited on the surface of a 4-inch LNOI wafer as etching hard mask utilizing plasma-enhanced chemical vapor deposition (PECVD). Various functional devices, including MZI, microring resonator and 3dB MMI are then patterned in the entire wafer using an ASML UV Stepper lithography system (NFF, HKUST) with a resolution of 500 nm. Next, the exposed resist patterns are transferred to the SiO$_2$ layer through a standard fluorine-based dry etching process, followed by transfer to the LN device layer using an optimized Ar+-based inductively coupled plasma (ICP) reactive-ion etching process. The etch depth for the LN layer is approximately 250 nm, leaving a 250-nm-thick slab. After removing the residual SiO$_2$ mask and redeposition, an annealing process is conducted. Subsequently, a second, third, and fourth lithography and lift-off processes are employed to fabricate the microwave electrodes, heater, and wires/pads, respectively. Finally, the chips are carefully cleaved for end-fire optical coupling.

**Principle of FM system**

The OFC generation is realized by the RF-driven double-pass PM, where the optical carrier is phase-modulated by an RF signal. Assuming a CW laser with a center frequency of $f_0$ and an RF driving signal with a frequency of $f_{LO}$, the optical field of the upper branch can be expressed as:

$$E_{upper}(t) = e^{-j2\pi f_0 t} \cdot e^{-j\beta \sin(2\pi f_{LO} t)} = e^{-j2\pi f_0 t} \cdot \sum_m J_m(\beta) e^{-j2\pi m f_{LO} t} \quad (1)$$

where $\beta$ is the modulation index, and $J_m(\beta)$ is the Bessel function of the first kind. By varying the input power of the RF signal, the amplitude $J_m(\cdot)$ of each optical sideband can be adjusted. From the perspective of the frequency domain, the output spectrum after phase modulation consists of a series of photonic harmonics with a spacing of $f_{LO}$ and a center frequency of $f_0$. An add-drop ring resonator is then employed to filter out specific photonic harmonics. The add-drop ring is designed to operate under over-coupled conditions to maximize the extinction ratio. It functions as a band-pass periodic filter with a frequency spectral range (FSR) of 196 GHz. By tuning the center wavelength of the add-drop ring, specific photonic harmonics can be selectively extracted. Finally, the output optical field of the upper branch could be expressed as $J_m(\beta)e^{-j2\pi(f_0+mf_{LO})t}$. The signal under test is modulated onto the optical carrier via a MZM using carrier-suppressed double-sideband modulation. Assume that the input RF signal with a frequency of $f_{RF}$, the output optical field of lower branch can be expressed as:

$$E_{lower}(t) = e^{-j2\pi f_0 t} \cdot \sin(\beta' \sin(2\pi f_{RF} t)) \quad (2)$$

where $\beta'$ denotes the modulation index. Under the small signal approximation ($\beta' \ll 1$), the optical higher-order sidebands can be ignored. Eq.(2) can be rewritten as:

$$E_{lower}(t) = \frac{\beta'}{2j}\left(e^{-j2\pi(f_0-f_{RF})t} - e^{-j2\pi(f_0+f_{RF})t}\right) \quad (3)$$

Finally, the upper and lower branches are combined through a 3 dB MMI coupler and sent to the PD, where the optical signal is converted into the electrical domain. The PD output $I(t)$ can be written as:

$$\begin{aligned} I(t) &= |E_{upper}(t) + E_{lower}(t)|^2 \\ &\propto \frac{\beta' J_m(\beta)}{2}[\cos(2\pi(f_{RF}+mf_{LO})t) - \cos(2\pi(f_{RF}-mf_{LO})t)] - \\ &\quad \frac{|\beta'|^2}{4}\cos(2\pi \cdot 2f_{RF} t) \end{aligned} \quad (4)$$

The DC output is neglected as it does not affect the final results. Generally, three frequencies are generated during the beating process. For the down-converter, the input frequency of the signals is greater than $m$ times the LO frequency ($f_{RF} > mf_{LO}$), resulting in a target output frequency of $f_{RF} - mf_{LO}$. The other spurious frequencies, $f_{RF} + mf_{LO}$ and $2f_{RF}$, fall outside the IF bandwidth and do not affect the in-band SSR. For the upconverter, there is not RF signal modulation. The signal is generated by beating the optical carrier and photonic harmonic $mf_{LO}$ directly, yielding a final generated frequency of $mf_{LO}$.

**Experimental methods and equipment**

The optical carrier from a tunable laser (Santec TSL-510) is amplified by an Erbium-doped fiber amplifier (EDFA). A polarization controller is then used to ensure transverse electric (TE) polarization, and the optical carrier is sent to the LN chip via a lensed fiber. The RF and LO signals are introduced to the chip through a high-speed GSGSG probe (GGB Industries, 50 GHz). The LO signal is generated by a signal generator (Agilent 83630B, 10 MHz - 26.5 GHz), while the RF signal is produced by a microwave source (Anritsu MG3697C, 2 GHz - 67 GHz). A DC bias voltage and RF signals are combined using a bias tee (SHF BT65R-D) to achieve carrier-suppressed double-sideband modulation, with the half-wave voltage of the MZM approximately 3 V. The output optical spectrum is recorded by an optical spectrum analyzer (YOKOGAWA AQ6370D). The output optical field is sent to a photodetector (XPDV412xR, 100 GHz) and recorded by an RF spectrum analyzer (Rohde & Schwarz FSW43, 10 Hz-43GHz). For the 110 GHz frequency measurement, we use an electrical down-converter (Shanghai AT Microwave M06HW, 110 GHz - 170 GHz) to reduce the center frequency, followed by two-stage electrical amplification with a gain of 35 dB.

**Funding.** Research Grants Council, University Grants Committee (CityU 11212721, CityU 11204022, C1002-22Y, N_CityU113/20); Croucher Foundation (9509005); City University of Hong Kong (9610682).

**Acknowledgement.** We thank the technical support of Mr. Chun Fai Yeung, Ms. Chan Olive, Mr. C W Lai, and Mr. Li Ho at HKUST, Nanosystem Fabrication Facility (NFF) for the stepper lithography and PECVD process. We thank Dr. Wing-Han Wong and Dr. Keeson Shum at CityU for their help in measurement and device fabrication.